\documentclass[twocolumn,aps,showpacs,pre,amsmath]{revtex4}
\usepackage{amssymb}
\usepackage{graphicx}

\begin{document}

\title{Interplay between Aharonov-Bohm interference and parity selective \\ tunneling in zigzag graphene nanoribbon rings}

\author{V. Hung Nguyen$^{1,2}$\footnote{E-mail: viethung.nguyen@cea.fr}, Y. Michel Niquet$^{1}$ and P. Dollfus$^3$}
\address{$^1$L$_-$Sim, SP2M, UMR-E CEA/UJF-Grenoble 1, INAC, 38054 Grenoble, France \\ $^2$Center for Computational Physics, Institute of Physics, Vietnam Academy of Science and Technology, P.O. Box 429 Bo Ho, 10000 Hanoi, Vietnam \\ $^3$Institut d'Electronique Fondamentale, UMR8622, CNRS, Universit$\acute{e}$ Paris Sud, 91405 Orsay, France}

\begin{abstract}
 We report a numerical study on Aharonov-Bohm (AB) effect and parity selective tunneling in $pn$ junctions based on zigzag graphene nanoribbon rings. We find that when applying a magnetic field to the ring, the AB interference can reverse the parity symmetry of incoming waves and hence can strongly modulate the parity selective transmission through the system. Therefore, the transmission between two states of different parity exhibits the AB oscillations with a $\pi-$phase shift, compared to the case of states of same parity. On this basis, it is shown that interesting effects such as giant (both positive and negative) magnetoresistance and strong negative differential conductance can be achieved in this structure. Our study thus presents a new property of the AB interference, which could be helpful to further understand the transport properties of graphene mesoscopic-systems.
\end{abstract}

\pacs{xx.xx.xx, yy.yy.yy, zz.zz.zz}

\maketitle

The Aharonov-Bohm (AB) interference \cite{ahbo59} is an elegant way of investigating phase coherent transport in mesoscopic systems. When a perpendicular magnetic-field $B$ is applied to a ring connected to two leads, the transmission probability exhibits oscillations with a characteristic period $\Delta B = \phi_0/S$, as consequence of a phase difference $\Delta \phi = 2\pi BS/\phi_0$ between the two arms forming the ring. Here, $S$ is the area of the ring and $\phi_0 = h/e$. The AB oscillations have been intensively studied and were observed in several systems such as metal rings \cite{webb85}, semiconductor heterostructures \cite{datt85}, carbon nanotubes \cite{bach99,lass07}, topological insulators \cite{peng10}, and recently in graphene rings (see the review \cite{sche12a}).

Due to its peculiar electronic properties, graphene has recently become a material of particular interest (e.g., see the review \cite{cast09}). Numerous unusual transport phenomena such as finite minimal conductivity, Klein tunneling, chiral properties of charge carriers, and unconventional quantum Hall effect have been explored in graphene nanostructures. Due to its specific electronic bandstructure, graphene has another remarkable property that the charge carriers in both conduction and valence bands can be involved simultaneously in the transport, as for instance in the Klein tunneling \cite{kats06} and band-to-band tunneling regimes \cite{vndo08}. This property has motivated studies on the AB effect in electron-hole graphene rings \cite{sche10,smir12,rahm13}, where a gate voltage was applied locally to one of two ring arms to generate interferences between electron-electron, hole-hole and electron-hole states. Theses studies allowed understanding of the interplay between the AB effect and Klein tunneling both away from and at the Dirac (charge neutrality) point. Additionally, they demonstrated the AB interference regardless of the charge carrier type in the ring arms \cite{rahm13}.
\begin{figure}[!t]
\centering
\includegraphics[width=3.0in]{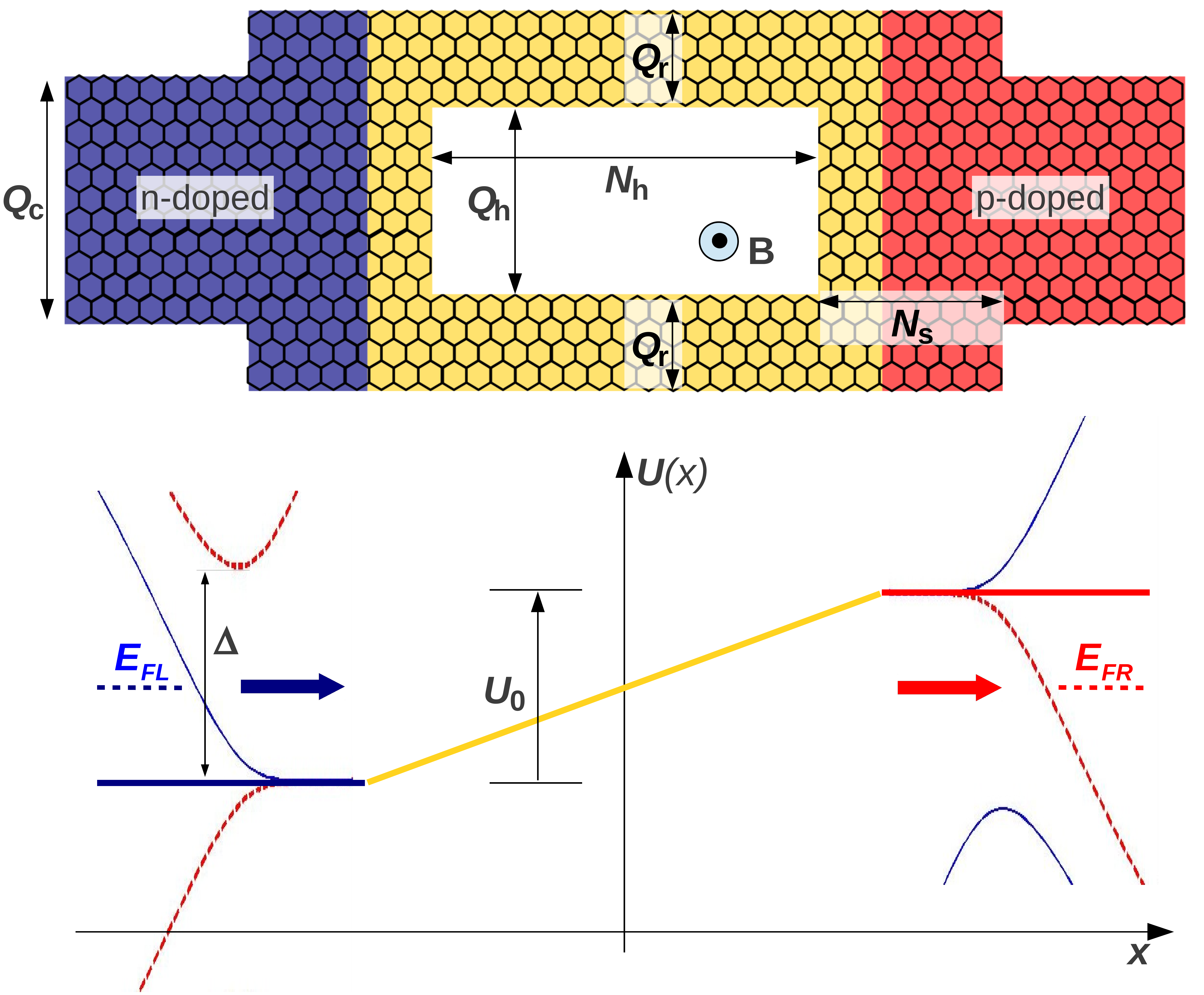}
\caption{Schematic of the graphene nanoribbon ring considered in this work and its energy band profile. $Q_r$, $Q_c$, and $Q_h$ characterize the width of the ring, of the contact graphene nanoribbons and of the hole, respectively. $N_h$ defines the length of the hole and $N_s$ stands for the size of side nanoribbons along the transport direction. $U_0$ is the potential difference between p-doped and n-doped zones.}
\label{fig_sim0}
\end{figure}

Besides the phenomena mentioned above, it has been shown that the lattice symmetry is a peculiar property of graphene nanoribbons (GNR). For instance, theoretical works have predicted that one-third of armchair edge GNRs are semimetals with a negligible bandgap and the rest are semiconductors while zigzag GNRs present spin polarized edges and edge localized states (see in ref. \cite{cast09}). Especially, in the zigzag GNR structures, some studies \cite{wang08,akhm08,ryce07,cres08,naka09} have predicted a phenomenon called the parity selective rule, which implies that the transmission between different subbands strongly depends on their even-odd parity. This phenomenon leads to interesting features such as the negative differential conductance (NDC) behavior, the valley-valve effect and the band selective filter. In a recent work \cite{hung13a}, we have demonstrated that strong AB oscillations can be achieved in rectangular GNR rings in the energy regime where only a single band is active. However, there is still an interesting subject, the interplay between the AB interference and parity selective tunneling, which has not been studied yet. In this article, we hence investigate the magnetotransport in $pn$ junctions based on zigzag GNR rings schematized in Fig. 1, where both effects can be observed simultaneously if the contact GNRs have an even number of zigzag lines. We will show that the parity selective rule can be modulated by the AB interference. In particular, the AB effect can reverse the parity symmetry of incoming waves, so that the transmission between states of different parity is significantly enhanced. We hence found a new property: compared to the case of same parity, there is a $\pi-$phase shift of AB oscillations when incoming and outgoing states have different parities. On this basis, strong AB oscillations with giant positive (resp. negative) magnetoresistance at low (resp. high) bias and strong NDC behavior can be achieved even at room temperature.

The transport properties of the GNR ring (see in Fig. 1) under a perpendicular magnetic field ($B-$field) are investigated using the Green's function method \cite{hung13b} to solve a nearest-neighbor tight-binding model as in ref. \cite{hung13a}. The presence of the $B-$field is included using the Peierls phase approximation \cite{peie33}. The hopping integral between nearest-neighbor atoms is hence given by $t_{nm} = -\tau_0 \exp(i\phi_{nm})$, where $\tau_0 \approx 2.7$ eV \cite{sche12b} and ${\phi_{nm}} = \frac{{2\pi }}{{{\phi _0}}}\mathop \smallint \nolimits_{{{\mathbf{r}}_n}}^{{{\mathbf{r}}_m}}\mathbf{A}(\mathbf{r})d\mathbf{r}$. The vector potential $\mathbf{A}(\mathbf{r}) = (-By,0,0)$ is related to the magnetic field $\mathbf{B} = (0,0,B)$ by $\nabla\times\mathbf{A} = \mathbf{B}$. The considered ring is characterized by the set of parameters of Fig. 1. The width of the GNRs (i.e., numbers of zigzag lines $Q_c$, $Q_r$, $Q_h$) is given in units of $3a_c/2$, their length ($N_h$, $N_s$) is given in units of $a_c\sqrt{3}$ with $a_c$ = 1.42 $\rm \AA$ and the difference between the potential energies in p-doped and n-doped zones is $U_0$. Within the Green's function framework, the transmission probability needed to evaluate the current and the local density of states (LDOS) are calculated as $\mathcal{T}\left( {\varepsilon, B} \right) = Tr\left[ {{\Gamma _L}{G^r}{\Gamma _R}{G^{r\dag }}} \right]$ and $D$($\varepsilon$,$B$;\textbf{r}$_n$) $=  - {\mathop{\rm Im}\nolimits} \left\{ {{G_{n,n}^r}\left( {\varepsilon, B} \right)} \right\}/\pi$, respectively, from the device retarded Green's function ${{G^r}\left( {\varepsilon, B} \right)}$, the injection rate ${\Gamma _{L\left( R \right)}} = i\left( {{\Sigma _L} - \Sigma _L^\dag } \right)$, and the self energy ${\Sigma _{L\left( R \right)}}$ defining the left (right) contact-to-device coupling. The current $\mathcal{I}\left( B \right)$ is computed using the Landauer formula \cite{hung13a} and the magnetoresistance (MR) is finally defined as $MR = \left[ {\mathcal{I}\left( B \right) - \mathcal{I}\left( 0 \right)} \right]/\mathcal{I}\left( B \right)$.

It has been shown in the literature that due to their lattice symmetry, each subband of zigzag GNRs has a parity characterizing the even-odd property of its wavefunction under a mirror reflection \cite{wang08,akhm08,ryce07,cres08,naka09}. This property leads to a parity selective rule on the transmission (i.e., parity selective tunneling mentioned above) through the GNRs having an even number of zigzag lines. In particular, the transmission between two states of different parity (e.g., as indicated in Fig.1) is forbidden while it is possible between states of same parity. We now go to investigate the AB interference for different parity properties of charge carriers. In Fig. 2, we display the transmission probability as a function of energy at different $B-$fields in a uniform zigzag GNR $pn$ junction and $pn$ junctions based on zigzag GNR rings. In all these cases, $Q_c$ is even, so that the parity selective tunneling manifests very well in the transport picture, i.e., at $B = 0$, a clear energy gap occurs consistently with the fact that the two states have a different parity while the ring is transparent beyond this gap.
\begin{figure}[!t]
\centering
\includegraphics[width=3.4in]{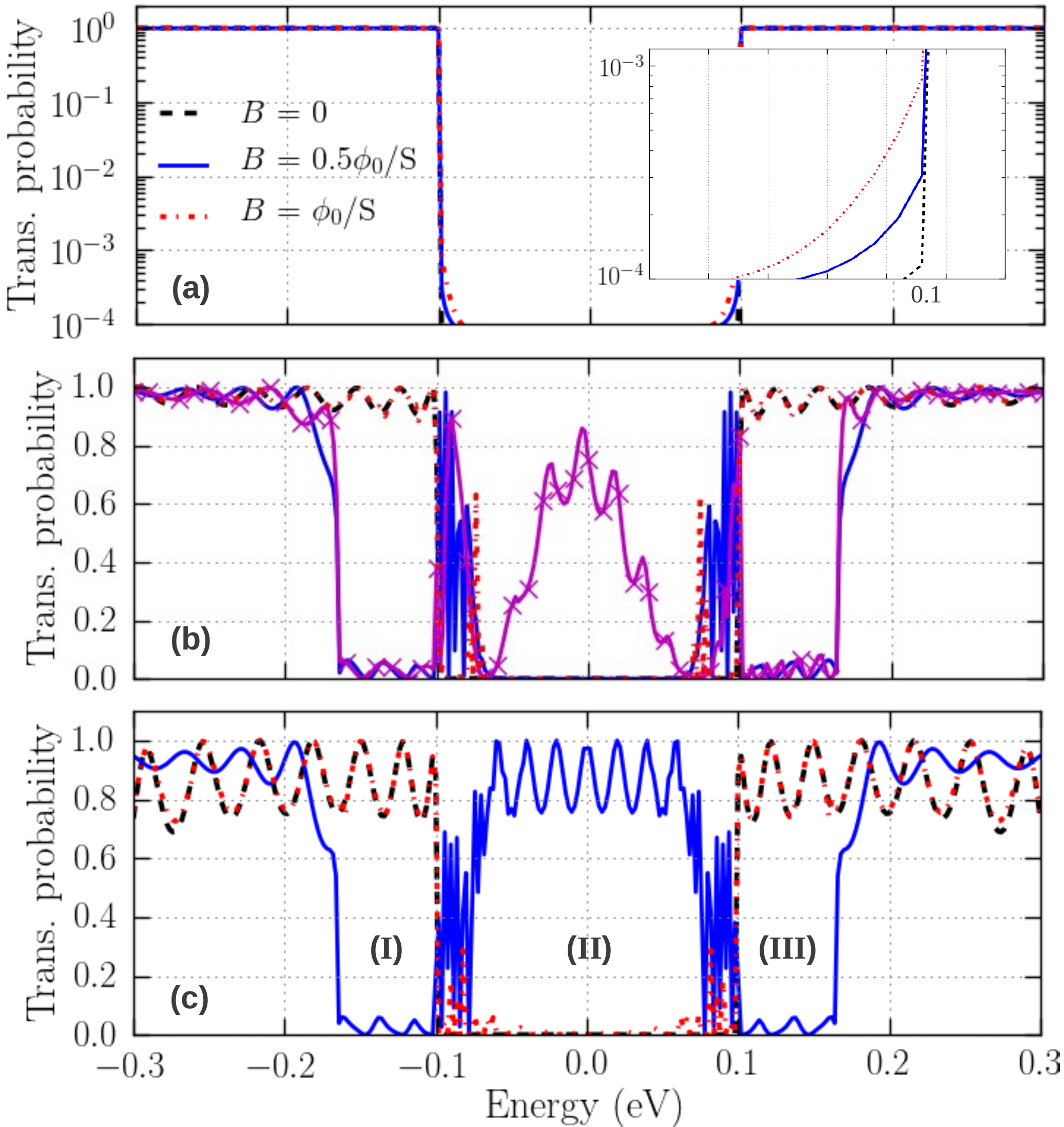}
\caption{Transmission probability as a function of energy at different $B-$fields: (a) uniform zigzag GNR $pn$ junction and (b,c) $pn$ junctions based on zigzag GNR rings with even ($= 22$) and odd $Q_r$ ($= 21$), respectively. The inset in (a) shows a zoomed image of the transmission probability around $E = 0.1$ eV. The $\times$-solid curve in (b) shows the data obtained at $B = 0.5 \times \phi_0/S$ in the GNR ring with $Q_r = 22$ where the the transition position between $n-$doped and $p-$doped zones is moved to the boundary between the left GNR contact and the ring. In (c) the energy regions (I), (II) and (III) distinguish different transport regimes (see in the text). $Q_c = 46$ everywhere and other parameters are $U_0 = 0.2$ eV, $Q_h = 14$ in (b) and 16 in (c), $N_h = 120$ and $N_s = 11$ in (b,c).}
\label{fig_sim1}
\end{figure}

First, let us introduce briefly the principle of this tunneling rule in the $pn$ junctions based on uniform zigzag GNRs. Based on the analysis in \cite{cres08}, any crystalline wave-function in the GNR made of $Q$ zigzag carbon chains can be expressed as a linear combination of Bloch sums
\begin{equation}
\Psi \left( \mathbf{k, r} \right) = \sum\limits_{j=1}\limits^{Q} {\alpha_j \Phi_j \left( \mathbf{k, r} \right) + \alpha_{2Q+1-j} \Phi_{2Q+1-j} \left( \mathbf{k, r} \right)}, \nonumber
\end{equation}
where $\Phi_j(\mathbf{k,r})$ is built from the 2$Q$ orbitals in the primitive cell (see eq. (3) of ref. \cite{cres08}) and $\mathbf{k}$ is the one-dimensional wave vector. In this equation, the first (resp. second) term is the lower (resp. upper) part of the wave-function while the coefficients $\alpha_i$ satisfy the relation $\alpha_{2Q+1-j} = \pm\alpha_j$ depending on its even/odd parity, which essentially arises from to the crystal symmetry of zigzag GNRs. The parity symmetry of each subband in two different GNR classes of even/odd number $Q$ has been shown in Figs. 5(͑a,b) of ref. \cite{cres08}. The transmission process illustrated in Fig. 1 was demonstrated to be possible only if the superimposed potential $U(x)$ acts as an inter-valley scattering source \cite{akhm08,cres08}. It was then shown in \cite{cres08} that the matrix element $U_{\alpha\beta}$ of $U$ between two states $\Psi_\alpha$(\textbf{k}) and $\Psi_\beta$(\textbf{q}) of different parity is zero in the GNRs of even number $Q$ and the transmission is hence blocked. Otherwise, $U_{\alpha\beta}$ is generally nonzero and the transmission is opened in the case of different parity or in the GNRs of odd $Q$. These features were named the parity selective tunneling \cite{wang08} and essentially explain the data of Fig. 2(a). The presence of a $B-$field, in principle, modifies the relation above between $\alpha_{2Q+1-j}$ and $\alpha_j$, so that the analysis above becomes no longer completely valid, i.e., the transmission probability in the gap region slightly increases when increasing the $B-$field (see the inset of Fig. 2(a)). However, in the range of $B-$field studied in this work, this effect is negligible, compared to the AB interference, as discussed below.
\begin{figure}[!t]
\centering
\includegraphics[width=3.4in]{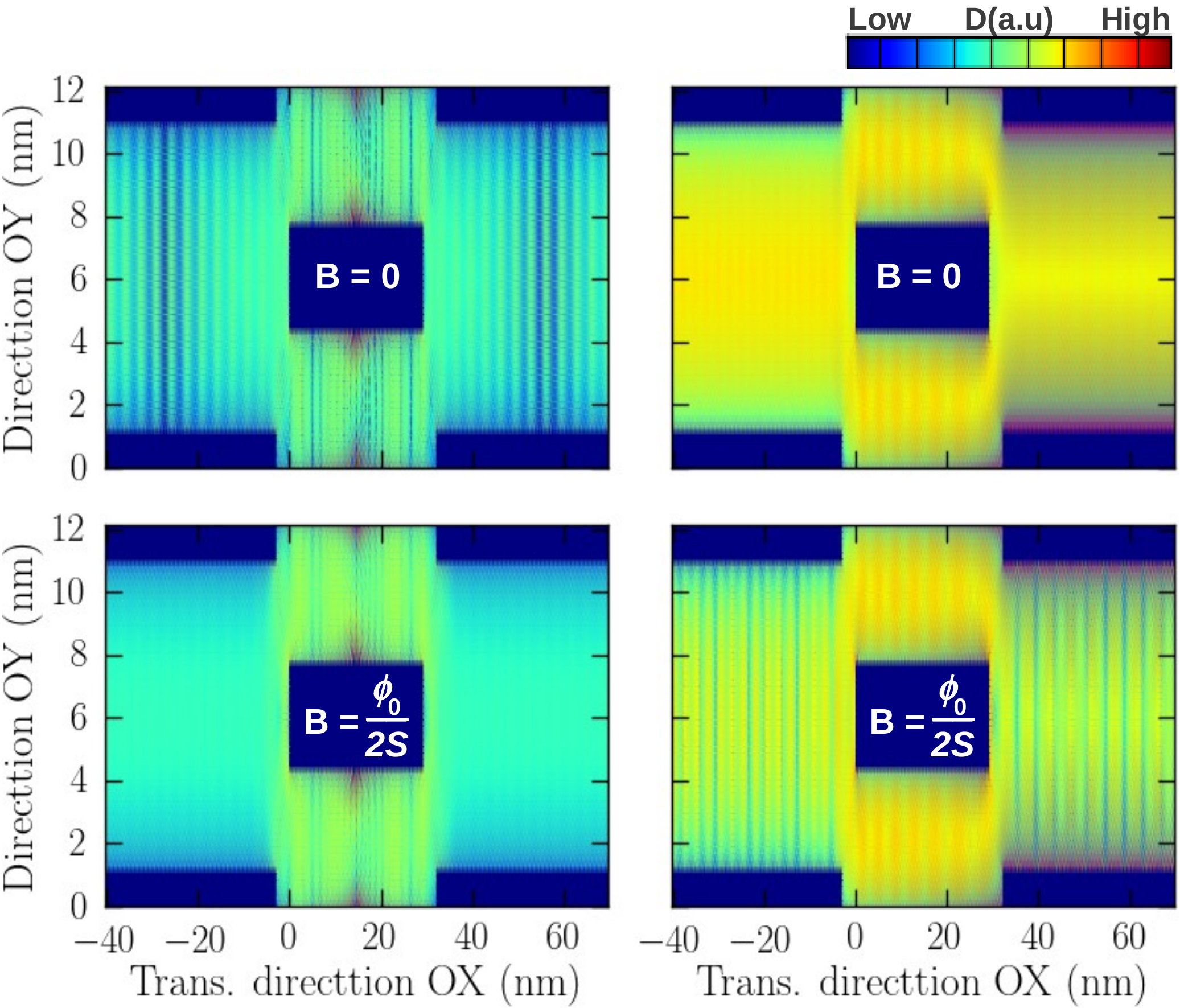}
\caption{Local density of states plotted at $E = 0$ (left panels) and 120 meV (right panels) for two different $B-$fields in the ring studied in Fig. 2(c).}
\label{fig_sim2}
\end{figure}

We now go to analyze the transport properties of $pn$ junctions based on zigzag GNR rings. At $B = 0$, the results displayed in Figs. 2(b,c) show that compared to the uniform GNR $np$ junction, the presence of the GNR ring does not affect the energy gap but results in resonant effects outside this gap. However, these resonant effects are not strong in these zigzag GNR rings, compared to that observed in amrchair GNR ones as discussed in \cite{hung13a}. Basically, the achievement of unchanged energy gap suggests that the parity selective rule is still valid in these considered heterostructures. When increasing the $B-$field, we find an interesting feature that the AB effect can change the parity selective tunneling, i.e., the transmission probability can be significantly enhanced in the gap regime. Moreover, the AB oscillations of the transmission between two states of different parity in the gap regime (II) have a $\pi-$phase shift compared to the case of same parity in the regimes (I) and (III), which will be shown more clearly by the $B-$dependence of the current later. A $\pi-$phase shift of the AB oscillations has been also observed in side-gated graphene rings \cite{huef10} by changing the gate voltages and explained by the generalized Onsager symmetry, which is essentially different from the phenomenon observed here, as discussed below. Note that due to the contribution of several subbands to the transport \cite{hung13a}, the AB effect is weak beyond the energy regimes (I-III). We also distinguish two different cases depending on the even or odd number $Q_r$ of the GNRs in the ring arms. The transmission exhibits strong oscillations in the whole gap region if $Q_r$ is odd (Fig. 2(c)). If $Q_r$ is even (Fig. 2(b)), these strong oscillations are just limited to the energies close to the potential energy in the n-doped and p-doped zones. In this case of even $Q_r$, we however have found as shown in Fig. 2(b) that if the transition position between n-doped and p-doped zones is moved outside of the ring arm to the boundary between the contact and the ring, the transmission can be significantly modulated in almost the whole gap region, similarly to the case of odd $Q_r$.

These obtained results can be explained as follows. The incoming wave when entering the ring is spatially separated into two parts. At zero $B-$field, these two parts when crossing the ring arms, because of the symmetry of two studied arms, still preserve the parity symmetry supported by the GNR contact. This explains why the parity selective rule is still valid in the considered rings at $B = 0$. However, in the presence of a finite $B-$field these two parts of the incoming wave are subjected to two different phase shifts \cite{ahbo59}, i.e., the upper (resp. lower) part is multiplied by a phase prefactor $\exp \left( { - i\frac{\phi(B)}{2}} \right)$ (resp. $\exp \left( { i\frac{\phi(B)}{2}} \right)$) with $\phi(B) = 2\pi BS/\phi_0$. Note that the ring area $S$ is determined as $S = (S_{inn}+S_{out})/2$ from the inner $S_{inn}$ and outer $S_{out}$ surfaces. In particular, $S \approx 258$ nm$^2$ and the period of AB oscillations $\Delta B = \phi_0/S \approx 16$ T here. These phase prefactors satisfy $\exp \left( { i\frac{\phi(B)}{2}} \right) = \exp \left( { - i\frac{\phi(B)}{2}} \right) = \left( -1 \right)^n$ if $B = n \phi_0/S$ while $\exp \left( { i\frac{\phi(B)}{2}} \right) = -\exp \left( { - i\frac{\phi(B)}{2}} \right) = \left( -1 \right)^n i$ when $B = \left( n + \frac{1}{2}\right) \phi_0/S$. The $B-$dependence of these two phase prefactors, in principle, yields the normal AB oscillations \cite{hung13a}, i.e., the transmission is blocked when $B = \left( n + \frac{1}{2}\right) \phi_0/S$, in the armchair GNR rings, the rings based on zigzag GNRs of odd number $Q$ and in the case of the transmission between two states of same parity (i.e., in the energy regimes (I) and (III)) in the rings considered here. However, the situation is opposite in the case of states of different parity shown above. In particular, because the two phase prefactors are identical when $B = n \phi_0/S$, the parity symmetry is still preserved as at $B=0$ and the transmission is forbidden. In contrast, these phase prefactors have opposite signs if $B = \left( n + \frac{1}{2}\right) \phi_0/S$, so that the parity symmetry of the incoming wave when crossing the ring is reversed (i.e., from odd to even or vice versa) to be the same as that of the outgoing wave, which opens the transmission. This phenomenon is further illustrated by Fig. 3 of the LDOS. In the energy regime (II) of Fig. 2, i.e., the case of different parity, the spatial variation of LDOS  (e.g., at $E$ = 0) along the transport direction is strong at $B = 0$ and the transmission is blocked, but it becomes invisible at $B = 0.5 \times \phi_0/S$, so that the structure is very transparent. On the opposite, in the regimes (I) and (III), i.e., the case of same parity, this variation  (e.g., at $E$ = 120 meV) is  invisible at $B = 0$ but strong at $B = 0.5 \times \phi_0/S$. Thus, the picture of phase shifts above explain well the transport phenomena observed.

The difference between the results displayed in Figs. 2(b,c) can be understood as follows. If $Q_r$ is even, the parity selective rule applies to the transmission along the ring arms, so that a clear energy gap is still observed. However, this effect of parity selective rule weakens and the transparence becomes high at the energy points close to the potential energy of the $n-$doped (resp. p-doped) zone because along the ring arms, the ratio between the GNR length, where the charge carriers are in the first conduction (resp. valence) band, and the length where they are in the first valence (resp. conduction) band is small. This effect vanishes if $Q_r$ is odd or if the transition point between n-doped and p-doped zones is moved outside of the ring arm to the boundary between the contact and the ring for even $Q_r$. The ring can hence be transparent in almost the whole gap region when applying a finite $B-$field.
\begin{figure}[!t]
\centering
\includegraphics[width=3.4in]{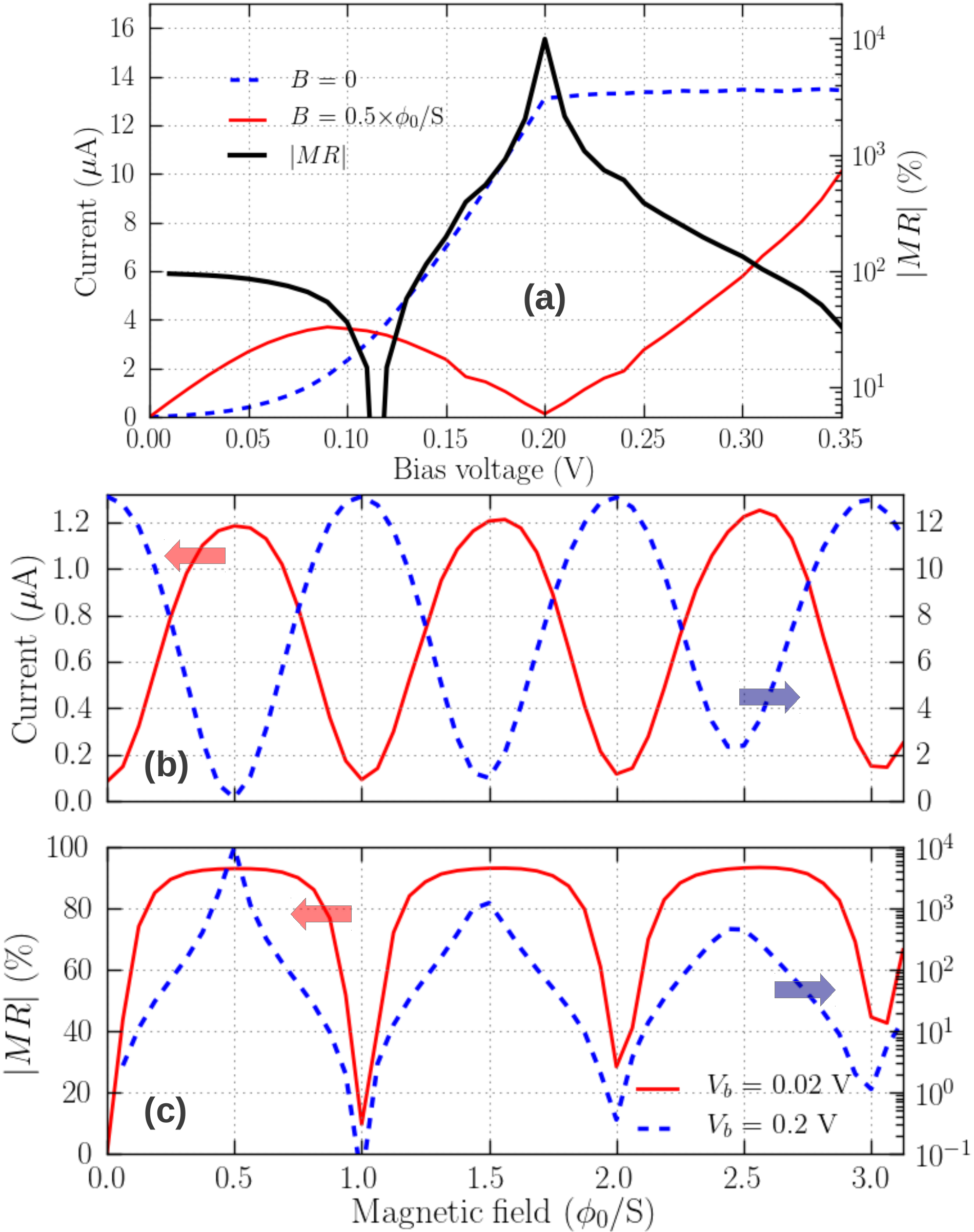}
\caption{(a) $I-V$ characteristics (see the left axis) of the GNR ring studied in Fig. 2(c) at $B = 0$ and $0.5 \times \phi_0/S$ and its corresponding MR (see the right axis). (b) Current and (c) MR as a function of $B-$field at different biases. All calculations were performed at room temperature.}
\label{fig_sim3}
\end{figure}

The transport properties observed are expected to govern interesting electrical behaviors of the ring. We display the current and corresponding magnetoresistance as a function of bias voltage in Fig. 4(a) and $B-$field in Figs. 4(b,c). At $B = 0$, the current is small at low bias due to the presence of the energy gap (II). When rising the bias, this gap reduces and hence the current increases. At high bias, when the potential profile ($U_R = U_0 - eV_b$) in the p-doped region moves down to come below that in the n-doped zone, this energy gap re-increases and the current reaches a saturation regime. Obviously, when varying the $B-$field, the current exhibits strong AB oscillations with a $\pi-$phase difference between low and high biases. Note that in the former case, the transmission in the energy regime (II) is dominant while in the latter one the regimes (I) and (III) govern the conduction. Accordingly, we observe (i) a giant positive (resp. negative) MR at low (resp. high) bias and (ii) a strong NDC behavior at $B = \left( n + \frac{1}{2}\right) \phi_0/S$, e.g., $B = 0.5 \times \phi_0/S$ here. In particular, a giant positive MR of about 95$\%$ at low bias, a giant negative MR (with a peak at $V_b = U_0/e$) of thousands $\%$ at high bias and the NDC behavior with a high peak-to-valley ratio (PVR) of about 28 are observed.

Interestingly, the rings studied here provide not only strong MR and NDC effects but also various possibilities to control these effects. In particular, we can modulate periodically the MR by tuning the $B-$field and can switch this quantity from a positive to a negative value by varying the bias. Similarly, the NDC behavior and the form of $I-V$ characteristics can be also modulated periodically by varying the $B-$field. The controllability of NDC behavior has been also observed by tuning the gate voltage in tunnel field-effect transistors (see ref. \cite{hung12} and references therein) and is suitable to design circuits operating at high frequency \cite{mizu95}. It is important to keep in mind that the NDC behavior observed here has a different origin from the similar effects previously explored in the literature, i.e., the peak (resp. valley) current is governed by the AB interference within the transmission between two states of different (resp. same) parity.
\begin{figure}[!t]
\centering
\includegraphics[width=3.4in]{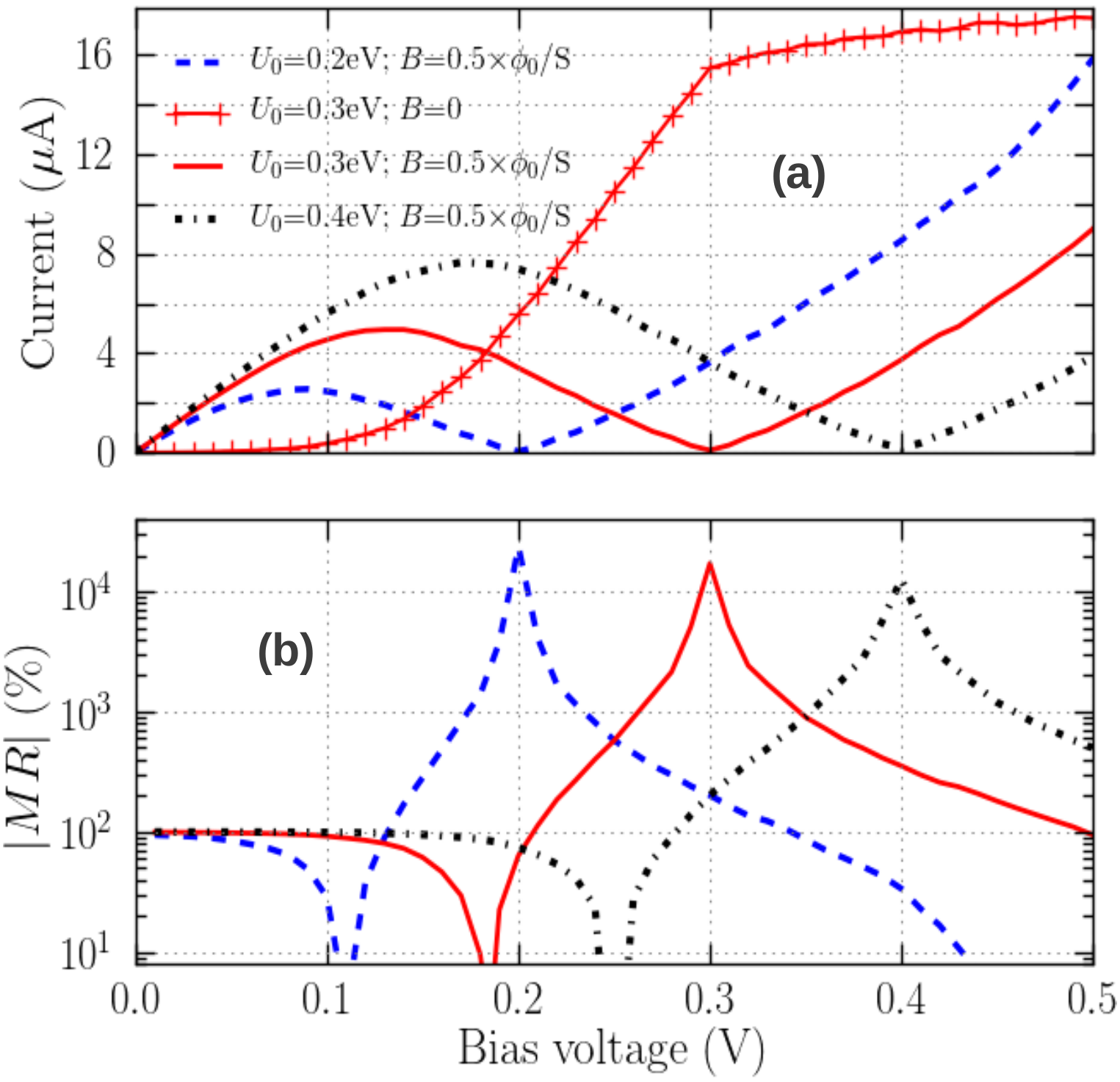}
\caption{(a) Current and (b) corresponding MR at $B = 0.5 \times \phi_0/S$ as a function of bias voltage in the GNR ring of $Q_c = 30$ with different $U_0$. Other parameters are the same as in the ring studied in Fig. 2(c). All calculations were performed at room temperature.}
\label{fig_sim4}
\end{figure}

We now would like to discuss the dependence of the MR and NDC effects on the structure parameters. Since the transport in the ring essentially depends on the balance between the contributions of the energy regimes (I-III) indicated in Fig. 2 (i.e., depends on $U_0$ and the bandstructure of contact GNRs), we display in Fig. 5 the $I-V$ characteristics and magnetoresistance at $B = 0.5 \times \phi_0/S$ obtained in the ring of smaller $Q_c = 30$ with different potential barriers $U_0$. Compared to the ring of $Q_c = 46$, the regimes (I) and (III) in this ring are larger, so that for the same $U_0 = 0.2$ eV, a larger MR peak ($\simeq$-2.4$\times10^4\%$ compared to -10$^4\%$) at $V_b = 0.2$ V and a stronger NDC behavior at $B = 0.5 \times \phi_0/S$ (with PVR of 62 compared to 28) are achieved. It thus suggests that we can enhance the value of the negative MR peak and the NDC behavior by reducing the width of contact GNRs. However, we also notice that as discussed in \cite{hung13a}, the condition $Q_c \gtrsim Q_h$ is mandatory to guarantee the strong AB effect in these rings. When increasing $U_0$, though the negative MR peak and the PVR are slightly reduced, we find that, because the energy regime (II) is enlarged, (i) the current peak of the NDC behavior at $B = 0.5 \times \phi_0/S$ increases significantly; (ii) the positive MR can approach almost 100$\%$; (iii) the bias windows of the giant positive and giant negative (i.e., $> 10^3\%$) MR are both enlarged. These properties are well pronounced in the range of $U_0 \lesssim 2\Delta$, where $\Delta$ (as illustrated in Fig. 1) is the energy spacing between the first and second band-edges. Beyond this range of $U_0$, the transport picture can be disturbed by the contribution of high-energy subbands, which has been shown to weaken the AB interference \cite{hung13a}.

Finally, we notice that to achieve the strong effects reported in this work, the good control of ribbon edges is necessary, which is really a technological challenge. However, on the one hand, the strong AB oscillations can still be achieved in the cases of weak edge-disorder, i.e., similarly to the results obtained for disorder probabilities $P_D < 10 \%$ in \cite{hung13a}. On the other hand, besides the top-down techniques successfully used to fabricate narrow GNRs at the nanometer scale, ultra-narrow GNR systems have been recently realized using surface-assisted bottom-up techniques \cite{jcai10,huan12,ruff12,blan12}, with atomically precise control of their topology and width. These techniques not only allow for the fabrication of ultra-narrow GNRs but also give access to GNR heterostructrures \cite{blan12}. Based on this, one can optimistically expect that the fabrication of the considered rings without or with a weak disorder can be achieved soon, which allow experimental verification of our predictions.

In summary, we have investigated the interplay between the AB interference and parity selective tunneling in $pn$ junctions based on zigzag GNR rings, where the contact GNRs have an even number of zigzag lines. We find that the AB interference can reverse the parity symmetry of incoming waves, so that the transmission between two states of different parity, which is blocked at $B = 0$, can be opened when applying a finite $B-$field. As an important result, the AB oscillations of these transmission processes exhibit a $\pi-$phase shift, compared to the case of states of same parity. On this basis, interesting effects such as giant (both positive and negative) MR and strong NDC behavior can be achieved at room temperature. The study also suggests possibilities to control/improve these effects by tuning the applied voltage, the $B-$field and the doping profile.

V.H.N. thanks Vietnam's National Foundation for Science and Technology Development (NAFOSTED) for financial support under Grant no. 103.02-2012.42. P.D. acknowledges the French ANR for financial support under the projects NANOSIM-GRAPHENE (Grant no. ANR-09-NANO-016) and MIGRAQUEL (Grant no. ANR-10-BLAN-0304).

\end{document}